\documentclass[article,twocolumn,showpacs,superscriptaddress]{revtex4-1}

\usepackage{graphicx}
\usepackage{amssymb}
\usepackage{bm}
\usepackage{braket}
\usepackage{amsmath}
\usepackage{tabularx}
\usepackage[dvipsnames]{xcolor}

\usepackage[colorlinks=true,
                linkcolor=blue,
	        citecolor=blue,
	        urlcolor=blue]{hyperref}

\newcommand\3{$_3$}
\newcommand{\theos}{Theory and Simulation of Materials (THEOS), and National Centre for Computational Design and Discovery of Novel Materials (MARVEL), \'Ecole Polytechnique F\'ed\'erale de Lausanne, CH-1015 Lausanne, Switzerland}
\newcommand{\dqmp}{Department of Quantum Matter Physics, University of Geneva, 24 Quai Ernest Ansermet, CH-1211 Geneva, Switzerland}

\AtBeginDocument{\usepackage{booktabs}}
\begin{document}
\title{Emergent dual topology in the three-dimensional Kane-Mele Pt$_2$HgSe$_3$}
\date{\today}
\author{Antimo Marrazzo}
\affiliation{\theos}
\author{Nicola Marzari}
\affiliation{\theos}
\author{Marco Gibertini}
\affiliation{\dqmp}

\begin{abstract}
Recently, the very first large-gap  Kane-Mele quantum spin Hall insulator was predicted to be monolayer jacutingaite (Pt$_2$HgSe$_3$), a naturally-occurring exfoliable mineral discovered in Brazil in 2008. The stacking of quantum spin Hall monolayers into a van-der-Waals layered crystal typically leads to a (0;001) weak topological phase, which does not protect the existence of surface states on the (001) surface. Unexpectedly, recent angle-resolved photoemission spectroscopy experiments revealed the presence of surface states dispersing over large areas of the 001-surface Brillouin zone of jacutingaite single crystals. The 001-surface states have been shown to be topologically protected by a mirror Chern number $C_M=-2$, associated with a nodal line gapped by spin-orbit interactions.
Here, we extend the two-dimensional Kane-Mele model to bulk jacutingaite and unveil the microscopic origin of the gapped nodal line and the emerging crystalline topological order.  By using maximally-localized Wannier functions, we identify a large non-trivial second nearest-layer hopping term that breaks the standard paradigm of weak topological insulators. Complemented by this term, the predictions of the Kane-Mele model are in remarkable agreement with recent experiments and first-principles simulations, providing  an appealing conceptual framework also relevant for other layered materials made of stacked honeycomb lattices. 
\end{abstract}

\maketitle
Graphene's crystal and electronic structure has been fundamental for the development of the theory of topological insulators. The very first model of a topological insulator ever proposed, namely the Chern (or quantum anomalous Hall) insulator by Haldane, is essentially a two-band tight-binding model for graphene in the presence of a staggered magnetic field with zero flux over the unit cell \cite{haldane_model_1988}.
The experimental isolation of graphene~\cite{novoselov_electric_2004} inspired Kane and Mele to assert that by doubling Haldane's model and introducing spins one could describe intrinsic spin-orbit coupling  (SOC) in graphene, leading to a novel gapped topological phase~\cite{kane_quantum_2005,kane_z2_05}. Such phase, identified by a  $\mathbb{Z}_2$ topological invariant, is named quantum spin Hall insulator (QSHI) and it is protected by time-reversal symmetry. Nowadays, graphene and the Kane-Mele (KM) model stand as one of the archetypal time-reversal invariant topological insulators, although negligible relativistic effects in carbon open only a vanishingly small band gap in graphene~\cite{bernevig_topological_2013}.

\begin{figure*}
\centering
\includegraphics[width = 1\linewidth]{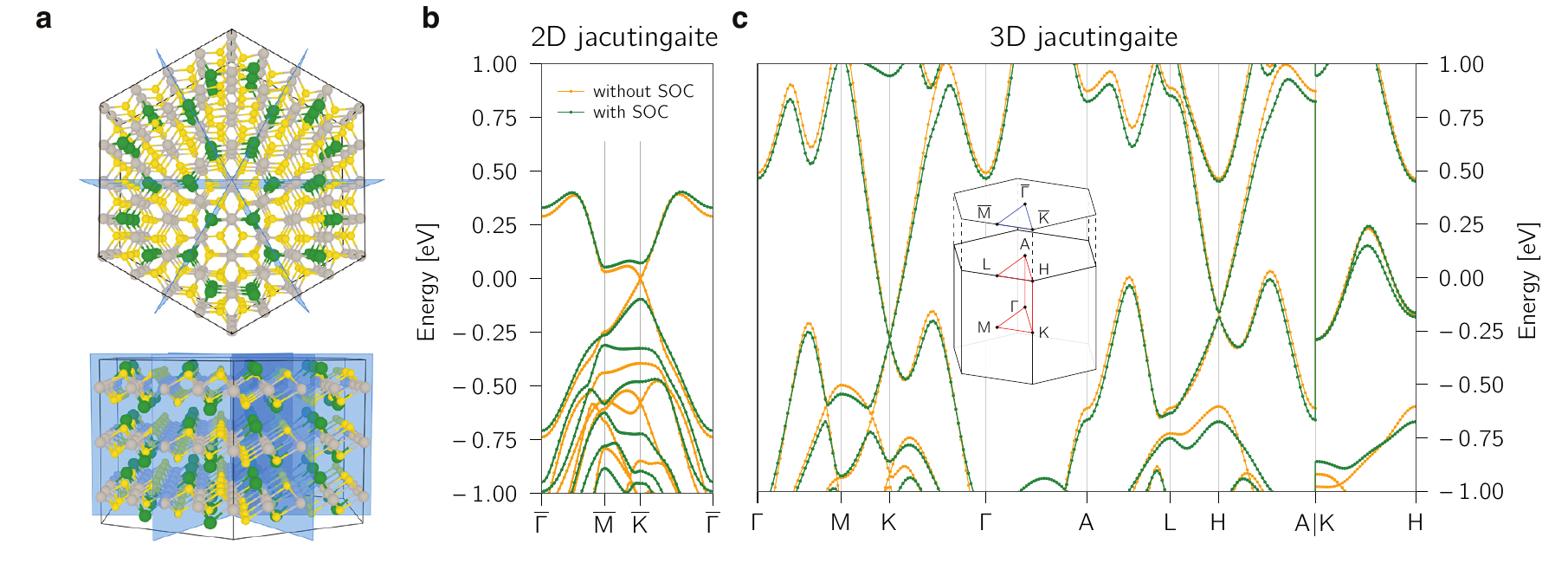}
   \caption{Crystal and electronic structure of jacutingaite. a) Top and lateral views of jacutingaite's crystal structure, where the three-fold symmetric mirror planes are highlighted in blue. Mercury, platinum, and selenium atoms are represented in green, grey, and yellow, respectively. b-c) Band structure of  monolayer (b) and bulk (c) jacutingaite along a high-symmetry path obtained using density-functional theory with (green) and without (orange) spin-orbit coupling (SOC). The inset shows the bulk and surface Brillouin zone. In monolayer jacutingaite the low-energy physics is described by graphene's Kane-Mele model (KM), with a substantial gap ($0.15$ eV with density-functional theory and $\sim0.5$ eV with many-body perturbation theory at the G$_0$W$_0$ level)  that opens at the Dirac point due to SOC. In 3D jacutingaite, Dirac cones are clearly visible at K and H and they are weakly gapped by  KM-type SOC; the band structure shows a strong dispersion along the stacking axis.}\label{fig1}
\end{figure*}

Notably, the KM model still applies to all Xenes~\cite{molle_buckled_2017}, i.e. the two-dimensional (2D) unary honeycomb materials made of group IV elements (e.g.\ silicene, germanene and stanene), where the presence of heavier atoms should lead to sizable band gaps driven by  KM SOC \cite{qshisilger_prl_2011,stanene_prl_2013,molle_buckled_2017}. Recently, monolayers of jacutingaite (Pt$_2$HgSe$_3$) have also been proposed as novel QSHIs that display the Kane-Mele physics, and at a much larger energy scale than in Xenes, with a band gap estimated to be around $\sim$ 0.5 eV \cite{Marrazzo2018}. Jacutingaite is a naturally-occurring layered mineral first discovered in 2008 \cite{cabral_first_obs_08} in a Brazilian mine and then synthesized in 2012 \cite{jacutingaite_exp_12}. Although jacutingaite is a ternary material with several differences with respect to the Xenes, it shares a (buckled) honeycomb structure of mercury atoms (see Fig.~\ref{fig1}a), which is ultimately responsible for the KM physics in monolayers~\cite{Marrazzo2018} that sparked experimental~\cite{Kandrai2019} and theoretical~\cite{Wu2019,facio_prm_2019,bansil_arxiv_2019,exp_jacu_2019} interest in this material.

Being a stacking of 2D QSHIs, bulk jacutingaite is expected to be a 3D weak topological insulator with indices $(0;001)$, and thus with no surface states on the $(001)$ surface~\cite{review_hasankane_2010}. Recent first-principles simulations confirmed this weak topological classification \cite{facio_prm_2019,bansil_arxiv_2019,exp_jacu_2019}, but at the same time surprisingly predicted the presence of basal surface states associated with a non-trivial mirror Chern number, thus promoting bulk jacutingaite to a dual topological material~\cite{Rauch2014,Eschbach2017} with both weak and crystalline topological properties. Such $(001)$ surface states have now been demonstrated independently through angle-resolved photoemission spectroscopy (ARPES) experiments on synthetic jacutingaite single crystals~\cite{exp_jacu_2019}. The unexpected dual topology of bulk jacutingaite cannot be understood through the standard paradigm of weak topological insulators \cite{review_hasankane_2010} and its interpretation opens interesting perspectives on non-trivial extensions of the KM model to describe 3D stacks of honeycomb layers.

In this Letter, we show that the non-trivial topology in bulk jacutingaite emerges from a strong interlayer hybridization that leads to a 3D generalization of the KM model including a large peculiar \emph{second nearest-layer hopping} term, while nearest layers are almost decoupled. Within this picture, even and odd layers are approximately independent and can be separately described by a 3D KM model where the novel hopping term drives a band inversion, giving rise to a nodal line that is gapped by SOC and a non-zero Chern number. Remarkably, when a coupling between even and odd layers is restored, the Chern numbers add up to a non-trivial value while the $\mathbb{Z}_2$ classification becomes weak, thus providing a microscopic understanding for the emergent dual topology of this material. 

We first shortly review the band structure of monolayer Pt\2HgSe\3 (shown in Fig.~\ref{fig1}b) and its relation to the KM model, as this will be instrumental to unveil the emergent topological properties of bulk jacutingaite crystals. When SOC is neglected, the valence and conduction bands touch at the Fermi energy, forming Dirac cones  located at the corners K/K$'$ of the 2D Brillouin zone (BZ). In a basis of maximally-localized Wannier functions~\cite{wannier_review_12} centered on mercury atoms~\cite{Marrazzo2018}, the linear dispersions arise from a hopping term between nearest neighbors on the (buckled) honeycomb Hg (sub)lattice, similarly to what happens in graphene. The inclusion of SOC  opens a substantial gap at K/K$'$, turning the system into a QSHI.  As discussed in Ref.~\onlinecite{Marrazzo2018}, gap opening mainly stems from a complex-valued second-nearest neighbor hopping, exactly as proposed by Kane and Mele, so that all qualitative features of the band structure of monolayer jacutingaite can be understood in terms of the KM model~\cite{kane_quantum_2005,kane_z2_05,liu_buckledSOC_11}:
\begin{align}\label{eq:KM}
H_{KM} = &-t \sum_{\langle ij \rangle\alpha}c_{i\alpha}^{\dagger}c_{j\alpha}+ i \Delta \sum_{\langle\langle ij \rangle\rangle\alpha\beta}v_{ij}s_{\alpha\beta}^z c_{i\alpha}^{\dagger}c_{j\beta} \notag\\
&+i \Delta' \sum_{\langle\langle ij \rangle\rangle\alpha\beta}u_{ij}({\bm s}\times {\bm d}_{ij}^0)_{\alpha\beta}^z c_{i\alpha}^{\dagger}c_{j\beta},
\end{align}
where the sums are restricted to pairs $\langle ij \rangle$ ($\langle\langle ij \rangle\rangle$) of first (second) nearest neighbor sites $i$ and $j$, $v_{ij}$, $u_{ij}$, ${\bm d}^0_{ij}$ are geometrical parameters~\footnote{$v_{ij}=\pm1$ depending on the orientation of the two nearest-neighbor bonds ${\bm d}_{1,2}$  (and can be written as ${\bm d}_1\times{\bm d}_2/|{\bm d}_1\times {\bm d}_2|$ ), $u_{ij}=\pm1$ for the two sublattices and ${\bm d}_{ij}^0$ is the unit vector connecting two second-nearest neighbors.}, and ${\bm s}= (s^x, s^y, s^z)$  are spin Pauli matrices. Here, in addition to the original KM hopping amplitudes $t$ and $\Delta$ associated respectively with the nearest-neighbor hopping and the KM SOC, we are adding an  ``in-plane'' SOC term with amplitude $\Delta'$, which is not present in planar honeycomb lattices such as graphene, but that appears when in-plane mirror symmetry is broken \cite{liu_buckledSOC_11} as is the case in monolayer jacutingaite.

Extending naively the analogy between graphene's KM model and monolayer jacutingaite to 3D, one would expect the electronic properties of bulk jacutingaite to be almost identical to its 2D form, as it is the case between graphene and graphite. In Fig.~\ref{fig1}c we report the band structure of bulk jacutingaite computed along a high-symmetry path by density-functional theory (DFT)~\cite{SM}. Without SOC, a linear dispersion is observed close to the K and H points, which is indeed reminiscent of the 2D Dirac cones. Still, the linear behavior extends over a much larger energy range than in 2D and a remarkable energy dispersion (compared to the overall band width) appears along the vertical direction between H and K. Even more compelling, SOC opens a gap between valence and conduction bands at K (and H) as in 2D, but the magnitude of the splitting is 1-2 orders of magnitude smaller than in the monolayer limit. Overall, these features suggest a significant coupling between layers, which is consistent with the non-negligible interlayer binding energy reported in Refs.~\onlinecite{mounet_nanotech_18,Marrazzo2018} ($\sim$ 60 meV$/$\AA$^2$, compared to 20 meV$/$\AA$^2$ for graphene) that sets jacutingaite as ``potentially exfoliable''~\cite{mounet_nanotech_18}.

 \begin{figure}
\centering
\includegraphics[width = \linewidth]{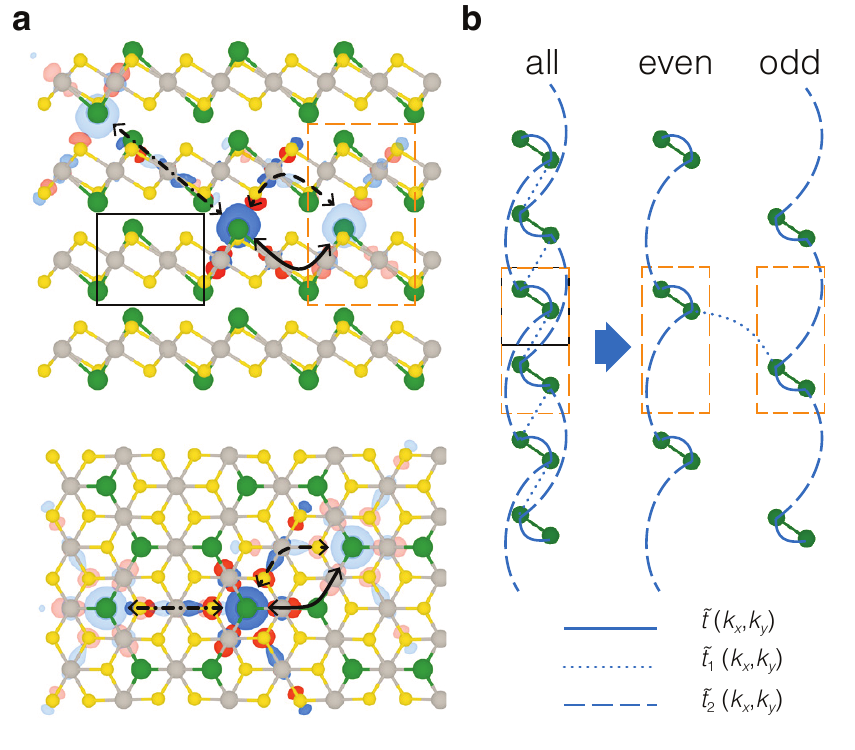}
    \caption{Wannier functions and even/odd layers decoupling. a) Top and lateral views of the maximally-localized Wannier functions (MLWFs) underlying the minimal model for bulk jacutingaite (see text). A reference MLWF is shown together with two additional ones (with lighter colors) to highlight the initial and final states of two relevant hopping terms in the model Hamiltonian: a second-nearest neighbor hopping process in the same layer (with two possible paths marked with solid or dashed lines) and a second-nearest layer hopping term (dash-dotted line) that drives bulk jacutingaite into a crystalline topological phase. b) Schematic of the most relevant hopping terms in the effective 1D model describing bulk jacutingaite at fixed parallel momentum $\mathbf{k}_\parallel=(k_x,k_y)$. The strongest terms are the intralayer $\tilde t(\mathbf{k}_\parallel)$ and second-nearest layer $\tilde t_2 (\mathbf{k}_\parallel)$, while $\tilde t_1(\mathbf{k}_\parallel)$ is almost negligible. This effectively leads to a decoupling between even and odd layers, which behave as a $\mathbf{k}_\parallel$-dependent Su-Schrieffer-Heeger-like chain with a doubled unit cell (orange dashed line) with respect to the primitive one (black solid line). 
    \label{FigMLWF}}
\end{figure}

To show that this significant interlayer coupling is responsible for the unexpected emergence of the bulk topology from the properties of the monolayer, we develop a minimal tight-binding model that captures all the relevant physics by extracting the most important hopping terms from the complexity of the full electronic structure. 
We first build a 4-band model (including spin) using Hg-centered maximally-localized Wannier functions (MLWFs) \cite{wannier_review_12} that reproduces the main features of the band structure around the Fermi level~\cite{SM}, such as the presence of SOC-gapped Dirac cones and the band dispersion between H and K. The corresponding Wannier functions are plotted in Fig.~\ref{FigMLWF}a: they clearly resemble the ones obtained for the monolayer in Ref.~\cite{Marrazzo2018}, but a notable difference is due to the spatial extension of the MLWFs; those of bulk jacutingaite spread over a neighboring layer, so that  they are effectively localized on 2 layers.

We now examine the strength of the hopping terms that appear in the MLWF Hamiltonian. The strongest term is the in-plane nearest-neighbor hopping $t$ that is responsible for the linear dispersion close to K and H, with a small splitting between valance and conduction bands which is opened by a very weak KM SOC. The fact that for bulk jacutingaite the effective KM SOC, as well as the in-plane SOC \cite{liu_buckledSOC_11,Marrazzo2018}, are strongly renormalized with respect to the monolayer can be understood by looking at the geometry of the MWLFs. Indeed, as shown in Fig.~\ref{FigMLWF}a, in bulk Pt\2HgSe\3 the overlap between MLWFs allows two alternative paths to hop to second-nearest neighbors within the same layer: one identical to monolayer jacutingaite (solid line) and one extending through the closest layer (dashed line). Owing to the different sign of the geometrical parameters $v_{ij}$ and $u_{ij}$ (see Eq.~\eqref{eq:KM}), the two paths give opposite contributions and, being very similar in magnitude, result in a very weak KM (and in-plane) SOC compared to the monolayer.

Remarkably, the second strongest hopping, which is comparable in magnitude to $t$, is a \emph{second nearest-layer} hopping term $t_2$ that connects two MLWFs as shown in Fig.~\ref{FigMLWF}a (dash-dotted line). Although the MLWFs that are involved are relatively far apart, the strong hopping amplitude stems from the partial delocalization of the MLWFs over the neighboring layer, which gives rise to a large overlap between MLWF that are two layers apart through the intermediate layer. Owing to the $\bar3m$ symmetry and the geometrical arrangement of Wannier functions, the strong overlap takes place between a reference MLWF and three others located two layers above (or below) in the opposite sublattice, thus mimicking  nearest-neighbor hopping but with a doubled in-plane separation~\cite{SM}. As we are going to see, the inclusion of this single term in the KM model is sufficient to understand the appearance of the nodal line and the surface states.

 \begin{figure*}
\centering
  \includegraphics[width =\linewidth]{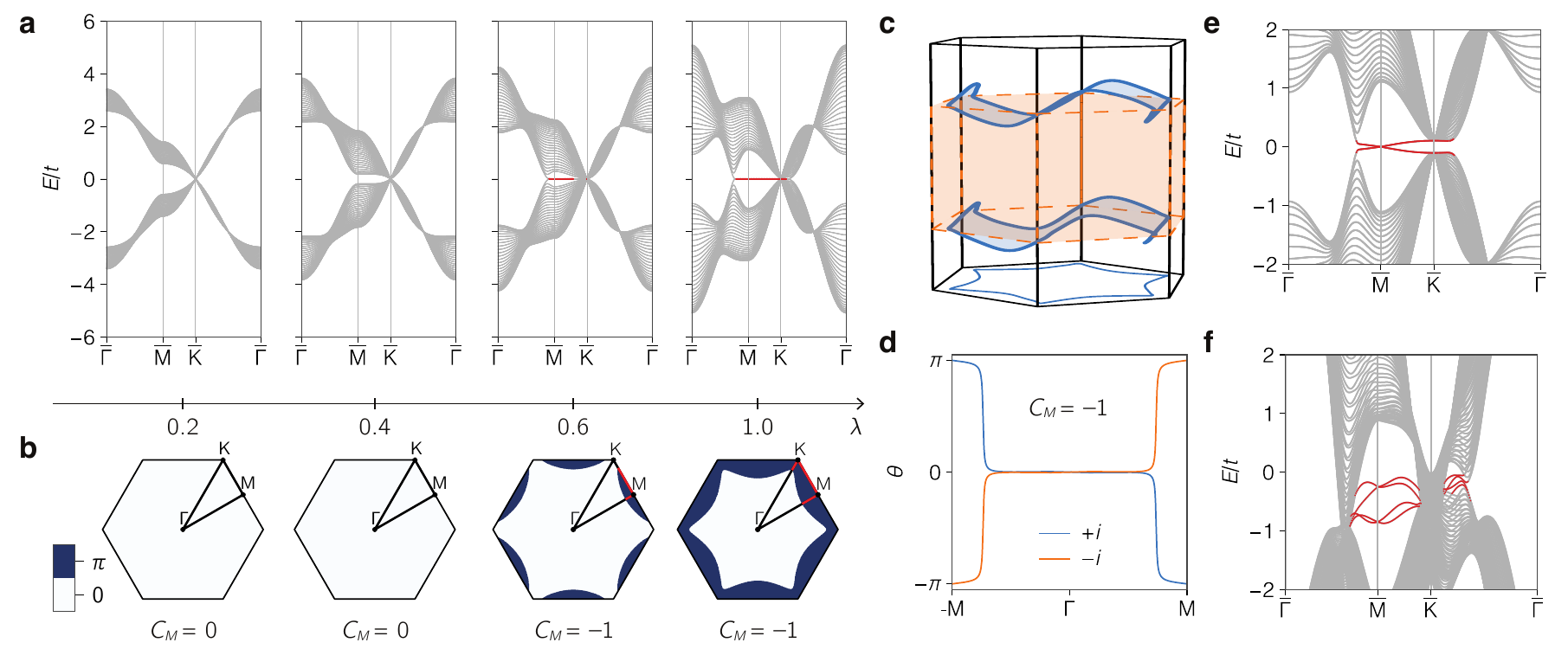}
    \caption{3D extended Kane-Mele model (J3KM) for jacutingaite.  a-b) Evolution of the J3KM model at zero SOC for different values of the interlayer coupling $\lambda$ in Eq.~\eqref{eq-model}: a) 30-layer slab band structure, with surface states  colored in red, and b) Zak phase \eqref{eq:Zak} over the 2D Brillouin zone (BZ). c) Nodal lines emerging in the J3KM model without SOC as a consequence of the second-nearest layer hopping term. The primitive BZ (black solid line) is shown together with the reduced one (orange dashed line) associated with the doubling of the unit cell. d-e) J3KM model with SOC: d) calculation of the mirror Chern number $C_M$ as the winding of the Wilson loop phase $\theta$ (restricted to the states with a definite $\pm i$ mirror eigenvalue) as a function of the parallel momentum in a mirror-invariant plane~\cite{Yu2011}. The evolution of $C_M$ with $\lambda$ is also reported in panel b). e) 30-layer slab band structure as in panel a) for $\lambda=1$ but with SOC included. f) 60-layer slab band structure for the J3KM model with SOC, when the coupling between even and odd layers is restored and additional symmetry-breaking terms are included. Four surface states are visible, in close agreement with experiments and direct DFT calculations of Ref.~\cite{exp_jacu_2019}. Band energies have been rescaled by the nearest-neighbor hopping amplitude $t$ ($=0.27$~eV in bulk jacutingaite) by setting the following parameters $t_2/t = -0.7$ and  $\Delta/t=\Delta'/t=0.02$ in Eq.~\eqref{eq:KM} and \eqref{eq-model}~\cite{SM}.  \label{Fig3}}
\end{figure*}

After having identified the most important hopping terms from the MLWF Hamiltonian, we build the following tight-binding model:
\begin{equation}
H_{J3KM} = H_{KM} +\lambda \tilde{H}_{2^{nd}NL}
\label{eq-model}
\end{equation}
where $H_{KM}$ is the KM model of Eq.~\eqref{eq:KM}, $\tilde{H}_{2^{nd}NL}$ is the second nearest-layer hopping just mentioned, and $\lambda$ is a dimensionless coupling constant that interpolates between graphene or monolayer jacutingaite ($\lambda=0$) and bulk jacutingaite ($\lambda=1$). 
This model is  a generalization of the Kane-Mele model to 3D jacutingaite and hereafter we will call it \emph{J3KM model}. As shown in Fig.~\ref{FigMLWF}b, at this level even and odd layers are completely decoupled, so it is natural to choose a unit cell that is doubled along the stacking axis and to describe separately even and odd layers through Eq.~\eqref{eq-model}. In this way, $\tilde{H}_{2^{nd}NL}$ becomes a first-nearest layer term in the even/odd subspace, and the BZ is halved along $k_z$.

For the sake of simplicity, we start by considering the J3KM model without SOC, corresponding to an Hamiltonian with only two terms: first nearest-neighbor (as in graphene) and inter-layer hopping. The band structure for a 30-layer slab is reported in Fig.~\ref{Fig3}a for different values of $\lambda$, where states localized at the (001) surface are colored in red. 
For $\lambda=1$ (corresponding to bulk jacutingaite), in addition to the graphene-like Dirac states at K, such simple model accounts for the presence of additional linear crossings between valence and conduction bands  (e.g. at a low-symmetry point between $\overline{\Gamma}$ and $\mathrm{\overline{M}}$), in perfect agreement with current ARPES measurements~\cite{exp_jacu_2019}. Notably, the model also exhibits 001-surface states that roughly span the same BZ region as observed in experiments~\cite{exp_jacu_2019}, thus providing a remarkably realistic qualitative description of the system, despite its simplicity. 

An interesting feature of the J3KM model that helps to rationalize the presence of surface states is that, at a given parallel momentum $\mathbf{k}_\parallel = (k_x,k_y)$, it is equivalent to a $\mathbf{k}_{\parallel}$-dependent 1D tight-binding Hamiltonian analogous to the Su-Schrieffer-Heeger (SSH) one, where the alternating hopping energies, $\tilde t(\mathbf{k}_{\parallel})$ and $\tilde t_2(\mathbf{k}_{\parallel})$, between intralayer and interlayer neighboring sites of a Hg chain depend parametrically on $\mathbf{k}_{\parallel}$ (see Fig.~\ref{FigMLWF}b and \cite{SM}). The polarization of this effective 1D chain can be written as $P (\mathbf{k}_{\parallel}) =e \gamma(\mathbf{k}_{\parallel})/(2\pi)$ in terms of the Zak phase~\cite{zak_1989,dhv_book_2018}:
\begin{equation}\label{eq:Zak}
\gamma(\mathbf{k}_{\parallel}) = -i \oint \braket{u(\mathbf{k}_\parallel,k_z)|\partial_{k_z}u(\mathbf{k}_\parallel,k_z)} d k_{z},
\end{equation}
where  $u(\mathbf{k}_\parallel,k_z)$ is the periodic part of the occupied one-electron eigenstate of the J3KM model at zero SOC.  
The combination of time-reversal and inversion symmetry dictates that the Zak phase can only assume two topologically distinct values: $\gamma(\mathbf{k}_{\parallel}) =\pi$ and $\gamma(\mathbf{k}_{\parallel}) =0$, depending on the relative strength of $\tilde t(\mathbf{k}_{\parallel})$ and  $\tilde t_2(\mathbf{k}_{\parallel})$ as in the SSH model. According to the surface charge theorem~\cite{Vanderbilt1993,dhv_book_2018}, the chain has an end charge whenever $\gamma(\mathbf{k}_{\parallel}) =\pi$ (polarization $e/2$), while it is topologically trivial when $\gamma(\mathbf{k}_{\parallel})=0$. We thus expect surface states in the J3KM for all values of $\mathbf{k}_\parallel$ for which $\gamma(\mathbf{k}_{\parallel}) =\pi$. Indeed, as shown in Fig.~\ref{Fig3}a and b, surface states in bulk jacutingaite ($\lambda=1$) appear precisely in regions of the BZ where the Zak phase is non-trivial.

The analogy with the SSH model also reveals that, along the lines that separate topologically distinct regions of the BZ we need to have $|\tilde t(\mathbf{k}_{\parallel})|= |\tilde t_2(\mathbf{k}_{\parallel})|$, so that the gap closes for some value of $k_z$. In other words, those lines can be considered as projections on the $(k_x,k_y)$ plane of nodal lines where the gap between valence and conduction bands vanishes. Indeed, by computing the energy bands of the J3KM model without SOC we uncover the presence of a nodal line (see Fig.~\ref{Fig3}c) dispersing across the border of the reduced BZ --also predicted by first-principles simulations~\cite{bansil_arxiv_2019,exp_jacu_2019}-- whose projection on the parallel plane is consistent with the boundary between regions with topologically distinct Zak phases~\footnote{The relationship between nodal lines and Zak (or Berry) phases is a general feature of systems with time reversal and inversion  symmetry~\cite{Yu2015,Kim2015,Fang2015,Chan2016,Fang2016} that goes beyond the applicability of the  SSH analogy.}.

We can now understand the emergence of surface states and nodal lines in the J3KM model by studying the evolution as a function of the interlayer coupling $\lambda$. Fig.~\ref{Fig3}a and b show the slab band structure and the Zak phase computed at different values of $\lambda$. For small $\lambda$, the band structure is essentially graphene-like, with no surface states and a trivial Zak phase over the full 2D BZ. With increasing $\lambda$, the occupied and empty bands get closer and closer to each other, until a band inversion occurs at the time-reversal-invariant point L of the reduced BZ (corresponding to $\overline{\mathrm{M}}$ in 2D). The band inversion creates three inequivalent nodal lines, whose projections separate regions with different Zak phases. Correspondingly, surface states appear in the slab calculation wherever $\gamma(\mathbf{k}_{\parallel}) =\pi$. With a further increase in $\lambda$, the three nodal lines merge into a single one, as shown in Fig.~\ref{Fig3}c for bulk jacutingaite ($\lambda=1$). The interlayer coupling thus plays a crucial role in driving the essentially trivial electronic structure of weakly coupled layers into the rich physics of bulk jacutingaite, with surface states associated with a nodal line in the absence of SOC. 

We now include SOC to show the robustness of surface states and their topological protection within the J3KM model. First, we consider the KM SOC~\cite{kane_quantum_2005,kane_z2_05} only, which gaps the nodal line almost everywhere, but not on the intersection with the vertical plane containing the $\Gamma$-M line (and its $3$-fold rotation symmetric partners). The inclusion of also the in-plane SOC~\cite{liu_buckledSOC_11,molle_buckled_2017} fully gaps the residual Dirac points and the system becomes a topological crystalline insulator, as supported by calculations of the mirror Chern number (see Fig.~\ref{FigMLWF}d) providing $C_M=-1$. As shown in Fig.~\ref{Fig3}e, the non-trivial Chern number protects the presence of 001-surface states even when  SOC is included, with a Dirac-like dispersion close to the $\overline{\mathrm{M}}$ point. 

Further calculations of the strong $\mathbb{Z}_2$ invariant $\nu$ show that the J3KM model for the even/odd subspace actually describes a strong $\mathbb{Z}_2$ topological insulator, in agreement with the fact that $\nu \equiv C_M\ \mathrm{mod}\ 2$ in this space group.
When considering together even and odd layers,  the $\mathbb{Z}_2$ invariant of the two subspaces is summed and becomes trivial ($1+1\equiv0\ \mathrm{mod}\ 2$), while the mirror Chern numbers add up to $C_M=-2$. On one side, this means that the weak $\mathbb{Z}_2$ topology of bulk jacutingaite does not fit the standard paradigm of weakly coupled 2D QSHI, but it is intimately related to the even and non-zero Chern number and to the double band inversion (one for each even/odd subspace) driven by the strong interlayer coupling. On the other, we expect the mirror Chern number to protect the surface states also when both subspaces are considered together.

In order to support this conclusion and at the same time to provide further evidence that the above predictions are not related to some extra symmetries (e.g. particle-hole) of the simple J3KM model, we finally consider the full MLWF Hamiltonian, which includes in particular additional  terms to Eq.~\eqref{eq-model} that: (i) restore the coupling between even and odd layers; (ii) introduce a finite dispersion along the K-H line as in the first-principles results of Fig.~\ref{fig1}c; and (iii)  break particle-hole symmetry, making the system a compensated semimetal. Still, they do not affect the topological classification of bulk jacutingaite. The corresponding band structure for a 60-layer slab is reported in Fig.~\ref{Fig3}f. Consistently with $C_M=-2$, two pairs of surface states (degenerate at $\overline{\mathrm{M}}$) are present, slightly split by the coupling between even and odd layers. Remarkably, these bands are very similar to what is observed in ARPES experiments~\cite{exp_jacu_2019}.

In conclusion, we  provide a microscopic insight on how symmetry-protected topological order in layered jacutingaite emerges from a non-trivial coupling between Kane-Mele-type QSHI monolayers. The essential physical features can be captured by a simple generalization of the Kane-Mele model to account for interlayer hopping. This J3KM model predicts the presence of surface states and nodal lines gapped by spin-orbit interactions, in remarkable agreement with recent ARPES measurements and first-principles simulations, providing  an appealing strategy to break the standard paradigm of weak topological insulators that becomes relevant for all other layered materials made of stacked honeycomb lattices. 

\section*{Acknowledgements}

We sincerely acknowledge F.\ Baumberger, I.\ Cucchi, and A.\ Tamai for useful discussions. 
This work was supported by the NCCR MARVEL of the Swiss National Science Foundation. M.G.\ acknowledges support from the Swiss National Science Foundation through the Ambizione program. Simulation time was awarded by CSCS on Piz Daint (production projects s825 and s917) and by PRACE on Marconi at Cineca, Italy (project id.\ 2016163963).

%

\clearpage

\onecolumngrid
  
\section*{\LARGE\bfseries Supplemental Material}

\renewcommand\theequation{S\arabic{equation}}
\renewcommand\thefigure{S\arabic{figure}}
\renewcommand\thesection{S\arabic{section}}
\renewcommand\thesubsection{S\arabic{section}.\arabic{subsection}}
\setcounter{equation}{0}
\setcounter{figure}{0}
\setcounter{table}{0}
\setcounter{section}{0}

\newcommand\kp{{\bm k}_\parallel}

\section{First-principles simulations}

Density-functional theory calculations are performed with the Quantum ESPRESSO distribution \cite{giannozzi_quantum_2009,giannozzi_qe_2017}, Wannier functions are obtained using 
WANNIER90 \cite{mostofi_updated_2014}.
Structural optimization is performed by using the experimental lattice parameters as obtained from X-ray diffraction and relaxing the atomic coordinates with a non-local van der Waals functional, namely the vdW-DF2 functional \cite{lee_df2_09} with C09 exchange (DF2-C09)  \cite{cooper_c09_10}, and the SSSP precision pseudopotential library v1.0 \cite{prandini_precision_2018} with 100~Ry of wavefunction cutoff and a dual of 8.
Further calculations on the optimized crystal structure (band structures and Wannier functions) are performed using the PBE functional \cite{perdew_pbe_96} and ONCV \cite{hamann_oncv_13} scalar and fully relativistic pseudopotentials from the PseudoDojo library \cite{dojo_paper_18} with 80~Ry of wavefunction cutoff and a dual of 4.
All calculation are perfomed with $\mathbf{k}$-point density of 0.09~\AA$^{-1}$,  that corresponds to a $\mathbf{k}-$point grid of $12\times12\times14$, and Marzari-Vanderbilt smearing \cite{mv_smearing_99} of 0.015~Ry. Wannier functions are constructed from a $\mathbf{k}-$point grid of $6\times6\times6$. Part of the calculations are powered by the AiiDA \cite{pizzi_aiida_16} materials' informatics infrastructure.

\section{4-band model for bulk jacutingaite}

\begin{figure}[h]
\centering
\includegraphics[width = 0.6\linewidth]{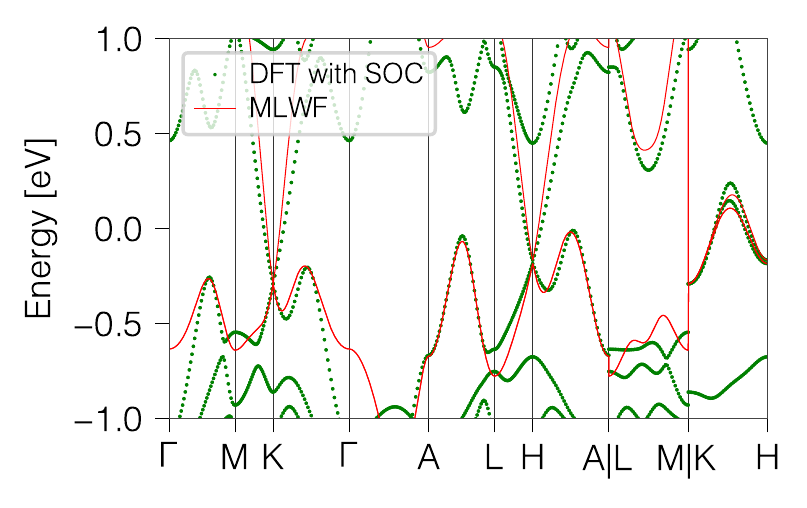}
\caption{Band structure of bulk jacutingaite along a high-symmetry path as obtained from first-principles calculations (green dots) and from the MLWF 4-band model (red line).  \label{fig:4band}}
\end{figure}

In order to construct a minimal tight-binding model for bulk jacutingaite, we follow a similar strategy as for the monolayer~\cite{Marrazzo2018} by mapping first-principles calculations onto a set of maximally-localized Wannier functions (MLWFs)~\cite{wannier_review_12}, constructed from an initial projection on Hg-centred $s$-like orbitals. This results in a 4-band (including spin and relativistic effects) tight-binding model on a lattice made of  buckled honeycomb layers directly stacked on top of each other (AA stacking). We denote the lattice vectors of the unit cell with ${\bm a}_1 = (a,0,0)$, ${\bm a}_2  = (-a/2,\sqrt{3}a/2,0)$, and ${\bm a}_3 = (0,0,c)$, with the two inequivalent sublattices centered at ${\bm \tau}_A = 2/3{\bm a_1} + 1/3{\bm a_2} + \delta {\bm a}_3$ and ${\bm \tau}_B = 1/3{\bm a_1} + 2/3{\bm a_2} - \delta {\bm a}_3$. The band structure obtained from this 4-band model is compared to the original first-principles results in Fig.~\ref{fig:4band}. Although a perfect quantitative agreement is not attainable, the  model reproduces all qualitative features around the Fermi energy, including in particular the extended linear dispersion and the very small band gap between valence and conduction bands at K and H. 

\section{Second-nearest-layer hopping term\label{sec:2ndNL}}

\begin{figure}[h]
\centering
\includegraphics{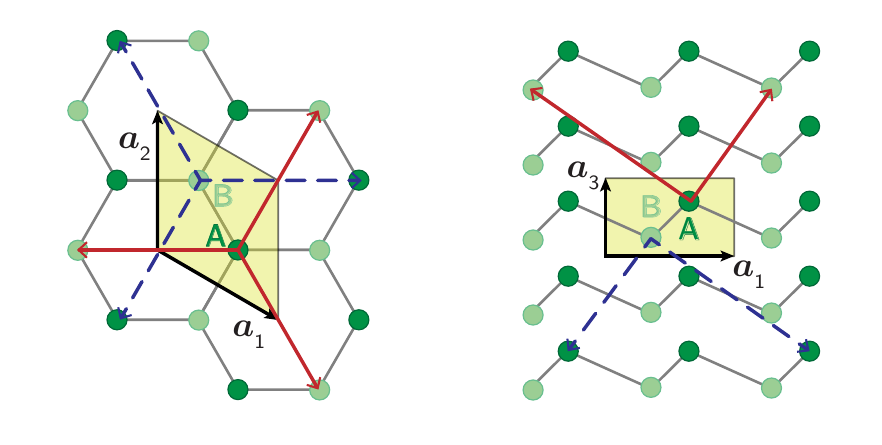}
\caption{Schematic representation of the buckled honeycomb lattice formed by Hg atoms on which the 4-band MLWF tight-binding model is defined. The A (B) sublattice is denoted with dark (light) green circles. The shaded yellow area shows the primitive unit cell, with lattice vectors ${\bm a}_1$, ${\bm a}_2$, and ${\bm a}_3$. Arrows connect sites involved in the second-nearest layer hopping process responsible for the non-trivial topological properties of bulk jacutingaite, starting from either the A sublattice (red solid arrows, corresponding to the vectors ${\bm \tau}_B+{\bm R}_{1,2,3}-{\bm\tau}_A$ with the upper sign in Eq.~\eqref{eq:2ndNLsep}) or the B sublattice (blue dashed arrows, corresponding to the vectors ${\bm \tau}_A+{\bm R}_{1,2,3}-{\bm\tau}_B$ with the lower sign in Eq.~\eqref{eq:2ndNLsep}).  \label{fig:sketch}}
\end{figure}

Within the 4-band model, the largest hopping term involves MLWFs that are centered on neighboring sites in the same layer, with amplitude $t=0.27$~eV. As mentioned in the main text, the next largest contribution is a second-nearest layer hopping term with amplitude  $t_2 = -0.18~\mathrm{eV}\simeq-0.7~t$. This hopping process involves a reference Wannier function on the A (B) sublattice and one of three B (A) sites that lie 2 layers above (below) in unit cells identified by the lattice vectors
\begin{equation}\label{eq:2ndNLsep}
{\bm R}_1 = \pm ( {\bm a}_1 - {\bm a}_2 + 2 {\bm a_3}),
\quad
{\bm R}_2 = \pm (-{\bm a}_1 - {\bm a}_2 + 2 {\bm a_3}),
\quad
{\bm R}_3 = \pm ( {\bm a}_1 + {\bm a}_2 + 2 {\bm a_3}).
\end{equation}
In Fig.~\ref{fig:sketch} we sketch the sites involved in this hopping process, starting either from the A sublattice (red solid arrows, upper signs in Eq.~\eqref{eq:2ndNLsep}) or the B sublattice (blue dashed arrows, lower signs in Eq.~\eqref{eq:2ndNLsep}).
This hopping term has been adopted in the main text to construct a minimal extension of the 2D Kane-Mele model that is able to describe the emergent topology in bulk jacutingaite. 

\section{Connection with the Su-Schrieffer-Heeger model}

In the absence of spin-orbit coupling, the J3KM model reduces to a 2-band model with a corresponding $2\times2$ Hamiltonian in Fourier space that can be written
\begin{equation}
{\mathcal H}(\kp,k_z) = 
\begin{pmatrix}\label{eq:fullSSH}
0  & t f(\kp) + t_2 g(\kp) e^{i2k_z c} \\
t f^*(\kp) + t_2 g^*(\kp) e^{-i2k_z c} & 0 
\end{pmatrix}
\end{equation}
where $\kp = (k_x,k_y)$ and according to Sec.~\ref{sec:2ndNL} we have 
\begin{align}
f(\kp) &= 1 + e^{i \kp\cdot {\bm a}_1}  + e^{-i \kp\cdot {\bm a}_2} = e^{i(k_x-\sqrt{3}k_y)a/2} \left[1 + 2e^{i \sqrt{3} k_y a/2} \cos\left(\frac{k_x a}{2}\right)\right] \\ 
g(\kp) &= e^{i \kp\cdot ({\bm a}_1+ {\bm a}_2)}  + e^{i \kp\cdot ({\bm a}_1 - {\bm a}_2)}  + e^{- i \kp\cdot ({\bm a}_1 +  {\bm a}_2)} = e^{i \kp\cdot ({\bm a}_1-{\bm a}_2)} f^*(2\kp) = e^{i(k_x-\sqrt{3}k_y)a/2} \left[e^{i \sqrt{3} k_y a} + 2 \cos(k_x a)\right]
\end{align}

If we now write $f(\kp) = |f(\kp)| e^{i\phi(\kp)}$, we thus have that $g(\kp) = |f(2\kp)| e^{-i\phi(2\kp) +i \kp\cdot ({\bm a}_1 -  {\bm a}_2)} $ and the Hamiltonian can be written as 
\begin{align}
{\mathcal H}(\kp,k_z) &= 
\begin{pmatrix}
0  & t |f(\kp)| e^{i\phi(\kp)} + t_2 |f(2\kp)| e^{i2k_z c-i\phi(2\kp) +i \kp\cdot ({\bm a}_1- {\bm a}_2)} \\
 t |f(\kp)| e^{-i\phi(\kp)} + t_2 |f(2\kp)| e^{-i2k_z c +i\phi(2\kp) -i \kp\cdot ({\bm a}_1- {\bm a}_2)}& 0 
\end{pmatrix}\notag\\
&= \begin{pmatrix}
1 & 0 \\
0 & e^{-i\phi(\kp)}
\end{pmatrix}
\begin{pmatrix}
0  & t |f(\kp)|  + t_2 |f(2\kp)| e^{2i [k_z - A(\kp)]c} \\
 t |f(\kp)|  + t_2 |f(2\kp)| e^{-2i [k_z - A(\kp)]c} & 0 
\end{pmatrix}
\begin{pmatrix}
1 & 0 \\
0 & e^{i\phi(\kp)}
\end{pmatrix}
\end{align}
where $A(\kp) = \left[\phi(\kp) + \phi(2\kp) - \kp\cdot ({\bm a}_1- {\bm a}_2)\right]/c $ is an effective vector potential. This is Hamiltonian is unitarily equivalent to the Hamiltonian of the Su-Schriffer-Heger (SSH) model in a magnetic field, where the hopping energies depend parametrically on the in-plane wave vector $\kp$ and the effect of the magnetic field is simply to shift the dispersion of the energy bands as a function of $k_z$ for a given value of $\kp$. The effective SSH hopping energies become 
\begin{equation}
\tilde t (\kp) = t | f(\kp)|
\qquad\text{and}\qquad
\tilde t_2(\kp) = t_2 |f(2\kp)|~.
\end{equation}
The condition for the gap to close reads $|\tilde t (\kp)| = |\tilde t_2 (\kp)|$, which has non-trivial solutions only when $|t_2/t|>1/3$. At the values of $\kp$ for which this condition is satisfied there must exist a value of $k_z$ at which the gap closes, thus giving rise to a nodal line. We have verified that, as expected, the projection of the nodal line on the $(k_x,k_y)$-plane (as identified by the solution of  $|\tilde t (\kp)| = |\tilde t_2 (\kp)|$ coincides with the line separating regions where the Zak phase (computed from the eigenstates of~\eqref{eq:fullSSH}) is trivial and regions where it is non-trivial. 

\end{document}